\documentclass[aps,prc,preprint,amsmath,amssymb,showpacs,preprintnumbers,superscriptaddress]{revtex4-1}
\usepackage{graphicx}
\usepackage{dcolumn}
\usepackage{bm}
\usepackage{mathrsfs}   
\usepackage{color,xcolor}  
\usepackage{multirow}  
\usepackage{booktabs} 

\usepackage{hyperref}

\newcommand{\la}{\langle}
\newcommand{\ra}{\rangle}
\newcommand{\beq}{\begin{equation}}
\newcommand{\eeq}{\end{equation}}

\newcommand{\bbm}{\begin{bmatrix}}
\newcommand{\ebm}{\end{bmatrix}}
\newcommand{\bBm}{\begin{Bmatrix}}
\newcommand{\eBm}{\end{Bmatrix}}
\newcommand{\bpm}{\begin{pmatrix}}
\newcommand{\epm}{\end{pmatrix}}

\begin{document}


\title{Nuclear matter in relativistic Brueckner-Hartree-Fock theory with Bonn potential in the full Dirac space}

\author{Sibo Wang}
\affiliation{State Key Laboratory of Nuclear Physics and Technology, School of Physics, Peking University \\
Beijing 100871, China}

\author{Qiang Zhao}
\affiliation{State Key Laboratory of Nuclear Physics and Technology, School of Physics, Peking University \\
Beijing 100871, China}

\author{Peter Ring}
\affiliation{Department of Physik, Technische Universit\"{a}t M\"{u}nchen, D-85747 Garching, Germany}

\author{Jie Meng}
\email{mengj@pku.edu.cn}
\affiliation{State Key Laboratory of Nuclear Physics and Technology, School of Physics, Peking University \\
Beijing 100871, China}
\affiliation{Yukawa Institute for Theoretical Physics, Kyoto University, Kyoto 606-8502, Japan}

\date{\today}

\begin{abstract}

Starting from the Bonn potential, relativistic Brueckner-Hartree-Fock (RBHF) equations are solved for nuclear matter in the full Dirac space, which provides a unique way to determine the single-particle potentials and avoids the approximations applied in the RBHF calculations in the Dirac space with positive-energy states (PESs) only. 
The uncertainties of the RBHF calculations in the Dirac space with PESs only are investigated, and the importance of the RBHF calculations in the full Dirac space is demonstrated.
In the RBHF calculations in the full Dirac space, the empirical saturation properties of symmetric nuclear matter are reproduced, and the obtained equation of state agrees with the results based on the relativistic Green's function approach up to the saturation density.

\end{abstract}



\maketitle


\section{Introduction}\label{SecI}

The nuclear \emph{ab initio} calculation, i.e., solving the nuclear many-body system starting from the bare nucleon-nucleon ($NN$) interaction, is one of the hot topics in nuclear physics~\cite{Baldo2007_JPG734-R243,Barrett2013_PPNP69-131,Hagen2014_RPP77-096302,Carlson2015_RMP87-1067,
Hegert2016_PR621-165,SHEN-SH2019_PPNP109-103713}.
Due to the strong repulsive core at short distance~\cite{Jastrow1951_PR81-165}, the bare $NN$ interaction cannot be directly applied within the conventional mean-field or Hartree-Fock (HF) approximation. Many methods including the Brueckner theory~\cite{Brueckner1954_PR95-217}, the low momentum $NN$ interaction $V_{\text{low}-k}$~\cite{Bogner2001_NPA684-432} and the similarity renormalization group (SRG)~\cite{Bogner2007_PRC75-061001} have been proposed to deal with the strong repulsive core. In the Brueckner theory, by summing all the ladder diagrams of the bare $NN$ interaction and taking into account the Pauli principle in the nuclear medium, an effective interaction, the $G$ matrix, is derived which incorporates the two-body short-range correlations induced by the strong repulsive core.

Replacing the bare $NN$ interaction by the $G$ matrix, the saturation properties of nuclear matter can be described qualitatively within the HF approximation~\cite{Day1967_RMP39-719}. However, in the non-relativistic framework, the saturation points of symmetric nuclear matter calculated by the Brueckner-Hartree-Fock (BHF) theory with two-body interactions are located on a so-called Coester line~\cite{Coester1970_PRC1-769}, which deviates systematically from the empirical values. The same is found by other non-relativistic \emph{ab initio} methods~\cite{Day1978_RMP3-495}. To solve this problem, it has been proposed to include three-body force (TBF) and the BHF calculations with TBF improve the description for the saturation properties of nuclear matter~\cite{LI-ZH2008_PRC77-034316,Vidana2009_PRC80-045806}. On the other hand, with two-body interactions only, the relativistic Brueckner-Hartree-Fock (RBHF) results~\cite{Brockmann1990_PRC42-1965} shift remarkably the saturation points close towards the empirical values, in contrast with those found in the non-relativistic BHF theory. This can be understood by the fact that, through virtual nucleon-antinucleon excitations in the intermediate states (the so-called $Z$ diagrams)~\cite{Brown1987CNPP}, relativistic effects lead to TBF.
RBHF theory has been widely applied to nuclear matter~\cite{LIU-L2002_ChPL19-190,MA-ZY2002_PRC66-024321,
VanDalen2004_NPA744-227,VanDalen2005_PRC72-065803,VanDalen2005_PRL95-022302,VanDalen2007_EPJA31-29,
Sammarruca2012_PRC86-054317,TONG-H2018_PRC98-054302,WANG-CC2020_JPG47-105108},
neutron stars~\cite{Engvik1994_PRL73-2650,Krastev2006_PRC74-025808,Katayama2013_PRC88-035805,TONG-H2020_PRC101-035802,
WANG-CC2020_ApJ897-96}, finite nuclei~\cite{SHEN-SH2016_CPL33-102103,SHEN-SH2017_PRC96-014316} and neutron drops~\cite{SHEN-SH2018_PLB778-344,SHEN-SH2018_PRC97-054312,
WANG-SB2019_PRC100-064319}.

The key point in the RBHF calculations for nuclear matter is to identify the single-particle potentials of the nucleons.
Due to the limitations of symmetries~\cite{Serot1986_ANP16-1}, the single-particle potential operator $\mathcal{U}$ is generally divided into the scalar and vector components. However, the effective interaction $G$ matrix has mixed the components through the solution of the scattering equation, and prevents a straightforward extraction of the single-particle potentials. Several methods have been proposed for the determination of single-particle potentials from the $G$ matrix, including the momentum-independence approximation~\cite{Brockmann1990_PRC42-1965}, the projection methods~\cite{Horowitz1987_NPA464-613,Gross-Boelting1999_NPA648-105}, the effective DBHF method~\cite{Schiller2001_EPJA11-15,MA-ZY2002_PRC66-024321}, and the solution of the scattering equation in the full Dirac space  including positive-energy states (PESs) and negative-energy states (NESs)~\cite{Anastasio1981_PRC23-2273,Poschenrieder1988_PRC38-471, Katayama2015_PLB747-43}.

As in the non-relativistic case, the full solution of the RBHF equations is an iterative process. 
Starting from an effective interaction $G^{(0)}$, which sums all the ladder diagrams with the Dirac spinors in free space, a single-particle potential ${\mathcal U}^{(0)}$ is determined. 
In the second step, the single-particle energies and Dirac spinors derived from this potential are used for the solution of the relativistic Bethe-Goldstone equation, and a new effective interaction $G^{(1)}$ is found. 
Based on this interaction, a new single-particle potential ${\mathcal U}^{(1)}$ is determined. 
This iteration goes on until the convergence is achieved.
In each iteration, for a unique determination of the single-particle potential ${\mathcal U}$, one needs the complete matrix elements of this operator, i.e. the matrix elements of $\mathcal U$ between PESs ($\mathcal U^{++}$), as well as those between PESs and NESs ($\mathcal U^{+-}$), and those between NESs ($\mathcal U^{--}$). 
The problem is, that one usually calculates the $G$ matrix with the similar codes for the calculation of $T$ matrix by the solution of the relativistic scattering equation, where only the scattering between nucleons, i.e. PESs, is considered. 
This means that in each iteration only $\mathcal U^{++}$ are well determined, which prevents the unique determination of ${\mathcal U}$.
Therefore in most of the RBHF calculations, ${\mathcal U}$ is calculated with various approximations.

The momentum-independence approximation~\cite{Brockmann1990_PRC42-1965} assumes that the single-particle potentials are independent of the momentum and the spacelike component of the vector potential is neglected. The scalar potential and the timelike component of the vector potential can be extracted directly from the single-particle potential energies at two selected momenta. However, the calculation suffers from uncertainties arising from the arbitrary choice for these two momenta. Moreover, this approximation fails to determine the correct behavior of the isospin dependence of the single-particle potentials~\cite{SHEN-H1997_PRC55-1211,Ulrych1997_PRC56-1788}.

In the projection methods, the $G$ matrix elements are projected onto a complete set of five Lorentz invariant amplitudes~\cite{Horowitz1987_NPA464-613}, from which the single-particle potentials are calculated analytically. However, the choice of these Lorentz invariant amplitudes is not unique. Different schemes of projections have been used~\cite{Horowitz1987_NPA464-613,Sehn1997_PRC56-216,Fuchs1998_PRC58-2022,Gross-Boelting1999_NPA648-105},
which differ mainly in the effect of the pseudoscalar meson exchange.

The effective DBHF method~\cite{Schiller2001_EPJA11-15,MA-ZY2002_PRC66-024321} suggests separating the $G$ matrix into the bare $NN$ interaction $V$ and a correlation term $\Delta G$. Then $\Delta G$ is parameterized in terms of a zero-range effective interaction with density-dependent coupling vertices. In this way, both the contributions of $V$ and $\Delta G$ to the single-particle potentials can be calculated within the relativistic Hartree-Fock (RHF) approach. Here the uncertainties originate from the different choice of the form and the parametrization of $\Delta G$.

As discussed above, the uncertainties in these methods are caused by the calculations in the Dirac space with PESs only, and can be cured by solving the RBHF equations in the full Dirac space, i.e., including PESs and NESs simultaneously~\cite{Nuppenau1989_NPA504-839,VanGiai2010_JPG37-064043}. In principle, the RBHF calculation for nuclear matter including NESs dates back to the 1980s. In the pioneering work of the Brooklyn group~\cite{Anastasio1980_PRL45-2096,Anastasio1981_PRC23-2273,Anastasio1981_PRC23-2258,Anastasio1983_PR100-327},
the relativistic effects were taken into account by expressing the single-particle wave functions for finite density by positive- and negative-energy spinors in free space and applying the first-order perturbation theory. Thus the $G$ matrix was calculated with the Dirac spinors in free space and the self-consistency was not achieved.

In the later of 1980s~\cite{Poschenrieder1988_PRC38-471}, a slightly different RBHF calculation in the full Dirac space was implemented utilizing the techniques of relativistic many-body Green's functions, in which the intermediate propagator in the scattering equation was chosen as the Brueckner propagator. In this way, Huber, Weber, and Weigel~\cite{Huber1995_PRC51-1790} found that the saturation properties of symmetric nuclear matter calculated with the Bonn potentials~\cite{Machleidt1989_ANP19-189} are in rather good agreement with the empirical values. However, it should be pointed out that the scattering equation for the $G$ matrix is different from the Thompson equation~\cite{Thompson1970_PRD1-110}, which is commonly used in most RBHF calculations.

In Ref.~\cite{Katayama2015_PLB747-43}, the RBHF calculation for nuclear matter was performed in the full Dirac space, where the Thompson equation was solved to obtain the $G$ matrix and the matrix elements of the bare $NN$ interaction are calculated in the rest frame of nuclear matter.
For symmetric nuclear matter an underestimation of the binding energy per nucleon about 5 MeV was found at saturation density for the potential Bonn A~\cite{Katayama2014_arXiv1410.7166}, compared to the empirical values of $-16\pm 1$\ MeV.

As manifested by the different predictions for the nuclear matter properties obtained in the literature~\cite{Huber1995_PRC51-1790,Katayama2015_PLB747-43}, fully self-consistent RBHF calculations in the full Dirac space are still an open problem~\cite{SHEN-SH2019_PPNP109-103713}.

In this work, to clarify the different predictions in the full Dirac space
~\cite{Huber1995_PRC51-1790,Katayama2015_PLB747-43}, we will perform the RBHF calculation including the PESs and NESs simultaneously. 
We choose the Thompson equation as the scattering equation and the matrix elements of the bare $NN$ interaction are treated in the c.m. frame. 
Apart from the inclusion of the NESs in each iteration, this scheme is consistent with the RBHF calculations in the Dirac space with PESs only. 

This paper is organized as follows. In Sec.~\ref{SecII}, the theoretical framework of the RBHF theory for nuclear matter in the full Dirac space is introduced. The numerical details are provided in Sec.~\ref{SecIII}. The calculated results and discussions are presented in Sec.~\ref{SecIV}. Finally, a summary is given in Sec.~\ref{SecV}.

\section{Theoretical framework} \label{SecII}

In the relativistic Brueckner-Hartree-Fock framework, the starting point is a bare $NN$ interaction in covariant form. In this work we adopt the one-boson-exchange interaction Bonn potential in Ref.~\cite{Machleidt1989_ANP19-189}, where the $NN$ interaction is mediated by the exchange of various bosons in terms of the following interaction Lagrangian densities coupling the meson fields to the nucleon
\beq\label{eq:Lagrangian}
  \begin{split}
    \mathscr{L}^{(pv)} =&\ -\frac{f_{ps}}{m_{ps}}\bar{\psi} \gamma^5 \gamma^\mu \psi \partial_\mu \varphi^{(ps)},\\
    \mathscr{L}^{(s)} =&\ g_s\bar{\psi} \psi \varphi^{(s)},\\
    \mathscr{L}^{(v)} =&\ - g_v\bar{\psi} \gamma^\mu \psi \varphi^{(v)}_\mu - \frac{f_v}{4M} \bar{\psi} \sigma^{\mu\nu} \psi
         \left(\partial_\mu \varphi^{(v)}_\nu - \partial_\nu  \varphi^{(v)}_\mu \right),
  \end{split}
\eeq
where $\psi$ denotes the nucleon field with the mass $M$. The bosons to be exchanged include the pseudoscalar $(ps)$ mesons $(\eta,\pi)$ with pseudovector $(pv)$ coupling, the scalar $(s)$ mesons $(\sigma,\delta)$, and the vector $(v)$ mesons $(\omega,\rho)$. For each pair, e.g., $(\eta,\pi)$, the first (second) meson has isoscalar (isovector) character. For isovector mesons, the field operator $\varphi_\alpha$ will be replaced by $\vec{\varphi}_\alpha\cdot\vec{\tau}$ with $\vec{\tau}$ being the usual Pauli matrices in isospin space, where the index $\alpha$ denotes different meson. $m_\alpha$ is the meson mass and the coupling strengths $f_\alpha$ and $g_\alpha$ are determined by fitting to the $NN$ scattering data and deuteron properties (see Ref.~\cite{Machleidt1989_ANP19-189}).

From the Lagrange density in Eq.~\eqref{eq:Lagrangian} the Hamiltonian density is obtained by the Legendre transformation. In the stationary case the Hamiltonian for nuclear system is found as an integral of the Hamiltonian density over the three-dimensional coordinate space and can be given in a second quantized form~\cite{Ring1996_PPNP37-193}
\beq
  H = \sum_{kl}   \la k|T|l\ra  b^\dag_k b_l + \frac{1}{2}\sum_{klmn} \la kl|V|mn\ra b^\dag_k b^\dag_l b_n b_m ,
\eeq
where the matrix elements are calculated as
\beq
  \begin{split}
    \la k|T|l\ra =&\ \int d^3r \bar{\psi}_k(\bm{r}) \left( -i\bm{\gamma}\cdot\bm{\nabla} + M \right) \psi_l(\bm{r}),\\
    \la kl|V|mn\ra  = &\ \sum_{\alpha} \int d^3r_1 d^3r_2 \bar{\psi}_k(\bm{r}_1) \bar{\psi}_l(\bm{r}_2) \Gamma_\alpha(1,2) D_\alpha(1,2)
     \psi_m(\bm{r}_1) \psi_n(\bm{r}_2).
  \end{split}
\eeq
Here $b^\dag_k$ and $b_k$ form a complete set of creation and annihilation operators for nucleons and the state $|k\ra$ stands for the Dirac spinor $\psi_k(\bm{r})$. The bare $NN$ interaction contains contributions from different mesons, with $\Gamma_\alpha$ and $D_\alpha$ being the interaction vertices and meson propagators, respectively.
The calculations of matrix elements $\la kl|V|mn\ra$ in momentum space are referred to Ref.~\cite{Brockmann1990_PRC42-1965} and Chapter 9 in Ref.~\cite{Weber1999book}. 
As mentioned in the introduction, in the Brueckner theory, the two-body short-range correlation induced by the strong repulsive core in the bare $NN$ interaction is incorporated into the $G$ matrix.

In the RBHF calculation, the $G$ matrix is obtained by solving the in-medium relativistic scattering equation, which is in strict analogy to the free-space scattering. The scattering equation in free space is chosen as the covariant Thompson equation~\cite{Thompson1970_PRD1-110}, which is one of the relativistic three-dimensional reductions of the Bethe-Salpeter equation~\cite{Salpeter1951_PR84-1232}. By replacing the Dirac spinors in free space by the ones in the nuclear medium and considering the Pauli principle, the Thompson equation is applied in the rest frame of nuclear matter in the form~\cite{Brockmann1990_PRC42-1965}
\beq\label{20190619-eq1}
  G(\bm{q}',\bm{q}|\bm{P},W)
  = V(\bm{q}',\bm{q}|\bm{P})
  + \int \frac{d^3k}{(2\pi)^3}
  V(\bm{q}',\bm{k}|\bm{P})
    \frac{M^{*}_{\bm{P}+\bm{k}}M^{*}_{\bm{P}-\bm{k}}}{E^*_{\bm{P}+\bm{k}}E^*_{ \bm{P}-\bm{k}}}
    \frac{Q(\bm{k},\bm{P})}{W-E_{ \bm{P}+\bm{k}}-E_{ \bm{P}-\bm{k}}}
  G(\bm{k},\bm{q}|\bm{P},W),
\eeq
where $E_{\bm{p}}$ is the eigenvalue of the Dirac equation in the nuclear medium (see Eq.~\eqref{20190819-eq1}).
$\bm{P}=\frac{1}{2}({\bm k}_1+{\bm k}_2)$ is the center-of-mass momentum and $\bm{k}=\frac{1}{2}({\bm k}_1-{\bm k}_2)$ is the relative momentum of the two interacting nucleons with momenta ${\bm k}_1$ and ${\bm k}_2$, and $\bm{q}, \bm{k}$ and $\bm{q}'$ are the initial, intermediate and final relative momenta of the two nucleons scattering in nuclear matter, respectively. The starting energy is denoted as $W$.
$M^*_{\bm{P}\pm\bm{k}}$ and $E^*_{\bm{P}\pm\bm{k}}$ are corresponding effective masses and energies (see Eq.~\eqref{20200819-eq12} and Eq.~\eqref{20190826-eq2} respectively). The Pauli operator $Q$ prohibits the scattering to the occupied states, i.e.,
\beq
  Q(\bm{k},\bm{P})=
  \begin{cases} 1,~&| \bm{P}+\bm{k}|,~|  \bm{P}-\bm{k}|> k_F\\
                0,~&\mbox{otherwise},
  \end{cases}
\eeq
with $k_F$ being the Fermi momentum.

Usually, Eq.~\eqref{20190619-eq1} is decomposed into partial waves in the helicity scheme~\cite{Jacob1959_APNY7-404} and reduced to an one-dimensional integral equation over the relative momentum $k$~\cite{Erkelenz1971_NPA176-413}
\beq\label{eqThomJ}
    \begin{split}
    \la \lambda'_1\lambda'_2|&G^J(q',q|P,W)|\lambda_1\lambda_2\ra
    =\la \lambda'_1\lambda'_2|V^J(q',q|P )|\lambda_1\lambda_2\ra
    +\sum_{h_1,h_2}\int \frac{k^2dk}{(2\pi)^3}
    \frac{M_{\text{av}}^{*2}(k,P)}{E_{\text{av}}^{*2}(k,P)}\times \\
    &\la \lambda'_1\lambda'_2|
    V^{J}(q',k|P )|h_1h_2\ra
    \frac{Q_{\mathrm{av}}(k,P)}{W-2E_{\text{av}}(k,P)}
    \la h_1h_2|G^J(k,q|P,W)|\lambda_1\lambda_2\ra,
  \end{split}
\eeq
where the indexes for the PESs and NESs have been suppressed for simplicity. $J$ stands for the total angular momentum for each partial wave. $\lambda_i, h_i$ and $\lambda'_i\ (i=1,2)$ denote the helicities of two nucleons in the initial, intermediate and final states.
To achieve this reduction, the Pauli operator $Q(\bm{k},\bm{P})$ is replaced by an angle-averaged Pauli operator $Q_{\mathrm{av}}(k,P)$~\cite{Erkelenz1974_PR13-191}, and the single-particle energies $E_{\text{av}}(k,P)$ and effective quantities $M^*_{\text{av}}(k,P)$ and $E^*_{\text{av}}(k,P)$ are calculated with the angle-averaged approximation $(\bm{P}\pm \bm{k})^2\approx \bm{P}^2 + \bm{k}^2$~\cite{Brockmann1990_PRC42-1965}.
In this work, the Thompson equation \eqref{eqThomJ} is solved in the full Dirac space by including PESs and NESs simultaneously for the initial and final states. For the intermediate states, the NESs are excluded due to the positive-energy projection operator in the Thompson propagator~\cite{Thompson1970_PRD1-110}.

In this work, the scattering equation \eqref{eqThomJ} is solved in the rest frame of nuclear matter. Since the bare $NN$ interaction, the Bonn potential, is determined in the two-body c.m. frame, a transformation for the matrix elements of the Bonn potential from the c.m. frame to the rest frame is necessary.
Usually one assumes that because of Lorentz invariance, the matrix elements of the Bonn potential in the rest frame are identical to those in the c.m. frame,  as in Ref.~\cite{Brockmann1990_PRC42-1965}.
However, the neglection of the retardation effects in the Bonn potential and the inclusion of one NES in the matrix elements will violate the Lorentz invariance of the matrix elements~\cite{Gross1992_PRC45-2094}.
The strict transformation from the c.m. frame to the rest frame is not trivial. 
For simplicity, we approximate the Lorentz invariance of the matrix elements of the bare $NN$ interaction in the following way
\beq\label{eq:transf}
  \la\lambda'_1\lambda'_2|V^J(q',q|P)|\lambda_1\lambda_2{\ra} \approx \la\lambda'_1\lambda'_2|V^J(q',q)|\lambda_1\lambda_2{\ra},
\eeq
i.e., the violation of the Lorentz invariance can be neglected.

The matrix elements of the bare $NN$ interaction in the c.m. frame for a given partial wave can be calculated as
\beq\label{Vmatrix}
  \la\lambda'_1\lambda'_2|V^J(q',q)|\lambda_1\lambda_2{\ra}=2\pi\int^{+1}_{-1} d(\cos\theta)d^J_{\lambda\lambda'}(\theta)\la \bm{q}'\lambda'_1\lambda'_2|V|\bm{q}\lambda_1\lambda_2{\ra},
\eeq
where $\lambda=\lambda_1-\lambda_2,\ \lambda'=\lambda'_1-\lambda'_2$. $\theta$ is the angle between $\bm{q}$ and $\bm{q}'$, and $d^J_{\lambda\lambda'}(\theta)$ are the conventional Wigner functions~\cite{Varshalovich1988}.
For the RBHF theory in the full Dirac space, both the PESs and NESs are included in the calculation in Eq.~\eqref{Vmatrix}.
With the approximation in Eq.~\eqref{eq:transf}, the Thompson equation \eqref{eqThomJ} in the full Dirac space can be solved using the partial-wave decomposition. Details are given in the Appendix \ref{App-ThEqu}.

In the RBHF theory, the nucleon inside the nuclear medium is regarded as a dressed particle in consequence of its interaction with surrounding nucleons. The single-particle motion in nuclear matter is described by the Dirac equation
\beq\label{20190819-eq1}
   (\bm{\alpha}\cdot\bm{p}+\beta M +\beta \mathcal{U} )u(\bm{p},s) = E_{\bm{p}}u(\bm{p},s),
\eeq
where $\bm{\alpha}$ and $\beta$ are the Dirac matrices and $u(\bm{p},s)$ is the positive-energy spinor with momentum $\bm{p}$, single-particle energy $E_{\bm{p}}$ and spin $s$. The medium effects are manifested by the single-particle potential (operator) $\mathcal{U}$. Due to the translational and rotational invariance, parity conservation, time-reversal invariance, and hermiticity in the rest frame of infinite nuclear matter, the single-particle potential $\mathcal{U}$ has the general form~\cite{Serot1986_ANP16-1}
\beq\label{20190610-eq7}
  \mathcal{U}(\bm{p}) = U_S(p)+ \gamma^0U_0(p) + \bm{\gamma\cdot\hat{p}}U_V(p),
\eeq
where $U_S(p), U_0(p), U_V(p)$\ are the scalar potential, timelike and spacelike components of the vector potential, respectively. $\hat{\bm{p}}=\bm{p}/|\bm{p}|$ is the unit vector parallel to the momentum $\bm{p}$.

With the definition of following effective quantities
\begin{align}
    \bm{p}^*=&\ \bm{p}+\hat{\bm{p}}U_V(p),\label{20200819-eq11}\\
    M^*_{\bm{p}}=&\ M+U_S(p),\label{20200819-eq12}\\
    E^*_{\bm{p}}=&\ E_{\bm{p}}-U_0(p)\label{20190826-eq2},
\end{align}
the Dirac equation in the nuclear medium can be expressed as
\beq
   \left(\bm{\alpha}\cdot\bm{p}^*+\beta M^*_{\bm{p}}\right) u(\bm{p},s) = E^*_{\bm{p}}u(\bm{p},s),
\eeq
where $E^*_{\bm{p}} = \sqrt{M_{\bm{p}}^{*2} + \bm{p}^{*2}}$. The positive-energy spinor $u$ and negative-energy spinor $v$ are obtained as
\begin{subequations}\label{eq:uv}
  \begin{align}
  u(\bm{p},s) =&\ \sqrt{\frac{E_{\bm{p}}^*+M_{\bm{p}}^*}{2M_{\bm{p}}^*}}
  		\bbm 1 \\ \frac{\bm{\sigma}\cdot\bm{p}^*}{E_{\bm{p}}^*+M_{\bm{p}}^*}\ebm \chi_s,
  		~~~~~~~~~~~~~~~~\qquad \bar{u}(\bm{p},s) u(\bm{p},s) = 1, \label{20200819-eq2} \\
  v(\bm{p},s) =&\ \gamma^5u(\bm{p},s) = \sqrt{\frac{E_{\bm{p}}^*+M_{\bm{p}}^*}{2M_{\bm{p}}^*}}
  		\bbm  \frac{\bm{\sigma}\cdot\bm{p}^*}{E_{\bm{p}}^*+M_{\bm{p}}^*} \\ 1  \ebm \chi_s,
  		\qquad \bar{v}(\bm{p},s) v(\bm{p},s) = -1, \label{20201207-eq1}
  \end{align}
\end{subequations}
where $\chi_s$ is the spin wave function.
The single-particle energies for PESs and NESs can be calculated as
\beq
  E^{+}_{\bm{p}} = E_{\bm{p}} = E^*_{\bm{p}} + U_0(p),\qquad E^{-}_{\bm{p}} = - E^*_{\bm{p}} + U_0(p).
\eeq

The Dirac equation can be solved analytically once the single-particle potentials are determined. To achieve this, three matrix elements of the single-particle potential operator $\mathcal{U} (\bm{p})$ are introduced as in Refs.~\cite{Anastasio1981_PRC23-2273,Poschenrieder1988_PRC38-471,Katayama2014_arXiv1410.7166},
\begin{subequations}\label{20200515-eq4}
  \begin{align}
    \Sigma^{++}(p)
    =&\   \bar{u}(\bm{p},1/2) \mathcal{U} (\bm{p}) u(\bm{p},1/2)
          = U_S(p)+ \frac{E^*_{\bm{p}}}{M^*_{\bm{p}}} U_0(p) + \frac{p^*}{M^*_{\bm{p}}} U_V(p), \label{20190829-eq1}\\
    \Sigma^{-+}(p)
    =&\  \bar{v}(\bm{p},1/2) \mathcal{U}(\bm{p}) u(\bm{p},1/2)
          = \frac{p^*}{M^*_{\bm{p}}} U_0(p)+ \frac{E^*_{\bm{p}}}{M^*_{\bm{p}}} U_V(p), \label{20190829-eq3}\\
    \Sigma^{--}(p)
    =&\   \bar{v}(\bm{p},1/2) \mathcal{U}(\bm{p}) v(\bm{p},1/2)
          = - U_S(p)+ \frac{E^*_{\bm{p}}}{M^*_{\bm{p}}} U_0(p) + \frac{p^*}{M^*_{\bm{p}}} U_V(p) , \label{20190829-eq2}
  \end{align}
\end{subequations}
where the direction of $\bm{p}$ is taken along the $z$-axis. 

Once $\Sigma^{++}, \Sigma^{-+}$ and $\Sigma^{--}$ are obtained, single-particle potentials can be determined uniquely through
\begin{subequations}\label{20191005-eq1}
  \begin{align}
    U_S(p) = &\ \frac{\Sigma^{++}(p)-\Sigma^{--}(p)}{2},\\
    U_0(p) = &\ \frac{E^*_{\bm{p}}}{M^*_{\bm{p}}}\frac{\Sigma^{++}(p)+\Sigma^{--}(p)}{2} - \frac{p^*}{M^*_{\bm{p}}}\Sigma^{-+}(p),\\
    U_V(p) = &\ -\frac{p^*}{M^*_{\bm{p}}}\frac{\Sigma^{++}(p)+\Sigma^{--}(p)}{2} + \frac{E^{*}_{\bm{p}}}{M^*_{\bm{p}}} \Sigma^{-+}(p).
  \end{align}
\end{subequations}

On the other hand, the three matrix elements of the single-particle potential operator in Eq.~\eqref{20200515-eq4} describe the single-particle potential energies of the nucleon with momentum $\bm{p}$. They can be calculated as the integrals over the effective interaction $G$ matrix:
\begin{subequations}\label{20200515-eq1}
  \begin{align}
    \Sigma^{++}(p) = &\ \sum_{s'} \int^{k_F}_0 \frac{d^3p'}{(2\pi)^3}\frac{M^*_{\bm{p}'}}{E^*_{\bm{p}'}}
    \la \bar{u}(\bm{p},1/2) \bar{u}(\bm{p}',s')| \bar{G}^{++++}(W)| u(\bm{p},1/2)u(\bm{p}',s')\ra, \label{20200218-eq3}\\
    \Sigma^{-+}(p) = &\ \sum_{s'} \int^{k_F}_0 \frac{d^3p'}{(2\pi)^3}\frac{M^*_{\bm{p}'}}{E^*_{\bm{p}'}}
    \la \bar{v}(\bm{p},1/2) \bar{u}(\bm{p}',s')| \bar{G}^{-+++}(W)| u(\bm{p},1/2)u(\bm{p}',s')\ra, \label{20200218-eq4}\\
    \Sigma^{--}(p) = &\ \sum_{s'} \int^{k_F}_0 \frac{d^3p'}{(2\pi)^3}\frac{M^*_{\bm{p}'}}{E^*_{\bm{p}'}}
    \la \bar{v}(\bm{p},1/2) \bar{u}(\bm{p}',s')| \bar{G}^{-+-+}(W)| v(\bm{p},1/2)u(\bm{p}',s')\ra. \label{20200218-eq5}
  \end{align}
\end{subequations}
$\bar{G}$ is the antisymmetrized $G$ matrix with the $\pm$-signs in the superscript denoting the PESs or NESs.
The \emph{no-sea} approximation~\cite{Walecka1974_APNY83-491} is used and thus the integrals are performed only for the single-particle states in the Fermi sea.
The factor $M^*_{\bm{p}'}/E^*_{\bm{p}'}$ is due to the fact that the Dirac spinors are normalized as, e.g., $\bar{u}u=1$ in Eq.~\eqref{20200819-eq2}.
In practice, the $G$ matrices in Eq.~\eqref{20200515-eq1} are calculated as a summation over different partial waves. Details are given in the Appendix \ref{App-Sgm}.

From Eq.~\eqref{20200515-eq4} to Eq.~\eqref{20200515-eq1}, it is clear that the full Dirac space provides a unique way to extract the single-particle potentials from the $G$ matrix, which effectively avoids the uncertainties of the calculations based on PESs only.

Eqs.~\eqref{eqThomJ}, \eqref{20200515-eq1}, \eqref{20191005-eq1} and \eqref{20190819-eq1} constitute a coupled set of equations that needs to be solved self-consistently. Starting from initial values of $U^{(0)}_S,U^{(0)}_0,U^{(0)}_V$ in vacuum, the Dirac spinors are obtained from the Dirac equation~\eqref{20190819-eq1}. Then one solves the Thompson equation \eqref{eqThomJ} to get the $G$ matrix and obtains $\Sigma^{++},\Sigma^{-+},\Sigma^{--}$ using the integrals in Eq.~\eqref{20200515-eq1}.
From Eq.~\eqref{20191005-eq1} a new set of values for $U^{(1)}_S,U^{(1)}_0,U^{(1)}_V$ are found to be used in the next iteration. This iterative procedure is repeated until a satisfactory convergence is reached.

Once the solution is converged, the binding energy per nucleon in nuclear matter can be calculated as
\beq\label{20210112-eq1}
  \begin{split}
  E/A
  =&\ \frac{1}{\rho} \sum_{s} \int^{k_F}_0 \frac{d^3p}{(2\pi)^3} \frac{M^*_{\bm{p}}}{E^*_{\bm{p}}}\la
    \bar{u}(\bm{p},s)| \bm{\gamma}\cdot\bm{p} + M |u(\bm{p},s)\ra - M \\
    &\ + \frac{1}{2\rho} \sum_{s,s'} \int^{k_F}_0 \frac{d^3p}{(2\pi)^3} \int^{k_F}_0 \frac{d^3p'}{(2\pi)^3}
     \frac{M^*_{\bm{p}}}{E^*_{\bm{p}}}\frac{M^*_{\bm{p}'}}{E^*_{\bm{p}'}}
     \la \bar{u}(\bm{p},s) \bar{u}(\bm{p}',s') |\bar{G}^{++++}(W)| u(\bm{p},s) u(\bm{p}',s') \ra,
  \end{split}
\eeq
where the isospin indices are suppressed. The starting energy $W=E_{\bm{p}}+E_{\bm{p}'}$. The density $\rho$ is related to the Fermi momentum $k_F$ through $\rho=2k^3_F/3\pi^2$. 
In parallel, the binding energy per nucleon can be calculated as the following as well
\beq\label{method2}
  E/A = \frac{1}{2\rho} \sum_{s} \int^{k_F}_0 \frac{d^3p}{(2\pi)^3} \left[ \frac{M^*_{\bm{p}}}{E^*_{\bm{p}}}\la 
    \bar{u}(\bm{p},s)| \bm{\gamma}\cdot\bm{p} + M |u(\bm{p},s)\ra + E_{\bm{p}} \right] - M.
\eeq
Eq.~\eqref{20210112-eq1} and Eq.~\eqref{method2} should lead to the same result.

The second derivate of $E/A$ with respect to the density $\rho$ at saturation density is the compression modulus $K_\infty$
\beq
  K_\infty = \left. 9\rho^2\frac{\partial^2 E/A(\rho)}{\partial \rho^2}\right|_{\rho=\rho_0},
\eeq
where $\rho_0$ is the saturation density.

\section{Numerical details}\label{SecIII}

In each iteration, by discretizing the momentum, the Thompson equation \eqref{eqThomJ} in the rest frame of nuclear matter leads to a set of matrix equations, which are solved by matrix inversion. 
The resulting $G$ matrix are used to determine the matrix elements of the single-particle operator (details are given in Appendix \ref{App-Sgm}) and a new set of single-particle energies $E_p$ and Dirac spinors with Eq.~\eqref{20190819-eq1}. 
They enter the Thompson equation \eqref{eqThomJ} in the next iteration.

This procedure depends crucially on the approximation introduced in the Lorentz transformation of the matrix elements of the bare $NN$ interaction in Eq.~\eqref{eq:transf}. 
Therefore it is very important to analyze the quality of this approximation and its influence on the binding energy. 
This can be done by comparing the results from two RHF calculations with the same effective interaction $V_{eff}$. One of these RHF calculations is carried out in the rest frame of nuclear matter and the other one in the c.m. frame. 
It is expected to get the same results, if we would carry out a proper Lorentz transformation between the two reference frames. 
The approximation in Eq.~\eqref{eq:transf} leads to two different results and by comparing the results we can check the quality of this approximation.
The calculation in the rest frame of nuclear matter is trivial. 
It corresponds to the solution of the conventional RHF equation as discussed in Ref.~\cite{Bouyssy1987_PRC36-380}. 
In this case the single-particle potentials are obtained by a variation of the energy functional with respect to the Dirac spinor, rather than using Eq.~\eqref{20191005-eq1}. 
The solution in the c.m. frame is more complicated because the calculation of the single-particle potentials with Eqs.~\eqref{20191005-eq1} and \eqref{20200515-eq1} requires the Lorentz transformation of four-vectors as discussed in Appendix \ref{App-Sgm}.

Comparing the results in the rest frame and in the c.m. frame with the effective force Wen $R(\delta)$ in Ref.~\cite{WEN-W2010_PTP123-811}, the deviation for the binding energy per nucleon is smaller than $0.2$ MeV in the density region $k_F=0.6 \sim 1.6\ \text{fm}^{-1}$. 
This shows the reliability of the approximation introduced in Eq.~\eqref{eq:transf}.

As usual, in (R)BHF theory, the single-particle potentials of the states with momentum $p$ above the Fermi momentum $k_F$ are not well defined. Different methods have been introduced in the literature, including the $gap\ choice$~\cite{Bethe1963_PR129-225}
and the $continuous\ choice$~\cite{Jeukenne1976_PR25-83}.
Here, a choice in between is adopted, where the single-particle potentials with momentum $p$ above $k_F$ are assigned to be equal to the ones at the Fermi momentum, i.e.,
\beq
  U_S(p) = U_S(k_F),\quad U_0(p) = U_0(k_F),\quad U_V(p) = U_V(k_F),\qquad \text{for}~~p>k_F.
\eeq
A similar treatment has been applied for finite nuclei in Refs.~\cite{Davies1969_PR177-1519,SHEN-SH2017_PRC96-014316}.

To calculate the matrix elements of the single-particle potential operator $\mathcal U$ from the $G$ matrix, the starting energies $W$ in Eq.~\eqref{20200218-eq3} - \eqref{20200218-eq5} need to be assigned. For $\Sigma^{++}$ in Eq.~\eqref{20200218-eq3}, one usually follows the Bethe-Brandow-Petschek (BBP) theorem~\cite{Bethe1963_PR129-225}
to choose $W=E^{+}_{\bm{p}}+E^{+}_{\bm{p}'}$. For $\Sigma^{-+}$ and $\Sigma^{--}$, one must clarify at first how to treat the NESs, either as occupied or as unoccupied states.
In the fully self-consistent RBHF calculation for finite nuclei in Ref.~\cite{SHEN-SH2018_PLB781-227},
it is found that, if the NESs are treated as occupied states, the ground-state properties for $^{16}$O are in better agreement with the experimental data. Moreover, the spin symmetry in the Dirac sea is better conserved in this choice~\cite{ZhouSG-2003PRL-91-262501,LiangHZ-2015PR-570-1,SHEN-SH2018_PLB781-227}.
Following Ref.~\cite{SHEN-SH2018_PLB781-227}, the NESs in Eqs.~\eqref{20200218-eq4} and \eqref{20200218-eq5} are treated as occupied states. For $\Sigma^{--}$ in Eq.~\eqref{20200218-eq5}, the starting energy is $W=E^{-}_{\bm{p}}+E^{+}_{\bm{p}'}$.
For $\Sigma^{-+}$ in Eq.~\eqref{20200218-eq4}, $\bar{G}^{-+++}(W)$ should be replaced by $\left[\bar{G}^{-+++}(W_1)+\bar{G}^{-+++}(W_2)\right]/2$, with $W_1 = E^{+}_{\bm{p}}+E^{+}_{\bm{p}'}$ and $W_2 = E^{-}_{\bm{p}}+E^{+}_{\bm{p}'}$.

In the calculation, the integrals over momentum and angle variables are discretized with $24$ and $12$ Gaussian grid points, respectively. The cutoff of total angular momentum is $10~\hbar$.
With these numerical conditions, the precision of binding energy per nucleon is less than $0.2$ MeV.

\section{Results and discussion}\label{SecIV}

\begin{figure}[htbp]
  \centering
  \includegraphics[width=8.0cm]{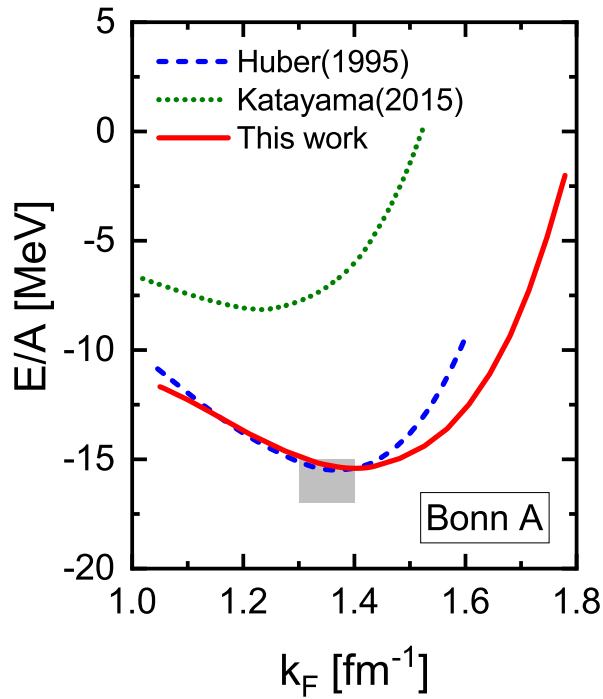}
  \caption{(Color online) Binding energy per nucleon $E/A$ of symmetric nuclear matter as a function of the Fermi momentum $k_F$ calculated by the RBHF theory in the full Dirac space with the potential Bonn A~\cite{Machleidt1989_ANP19-189}.
  Our result (red solid line) is compared with Ref.~\cite{Huber1995_PRC51-1790}
  (blue dashed line) and Refs.~\cite{Katayama2015_PLB747-43,Katayama2014_arXiv1410.7166}
  (green dotted line). The shaded area indicates the empirical values.}
  \label{Fig1}
\end{figure}

We perform the RBHF calculation in the full Dirac space with the bare $NN$ interaction chosen as the relativistic potential Bonn A~\cite{Machleidt1989_ANP19-189}, where the scattering equation is chosen as the Thompson equation~\cite{Thompson1970_PRD1-110} and the matrix elements of Bonn A are treated in the c.m. frame.
Fig.~\ref{Fig1} shows the binding energy per nucleon $E/A$ of symmetric nuclear matter as a function of the Fermi momentum $k_F$. The shaded area indicates the empirical values. It can be seen that the nuclear matter saturation point is reasonably described in this work, much better than in a non-relativistic calculation with Bonn A~\cite{Brockmann1990_PRC42-1965}, where the saturation energy is -23.55 MeV and the saturation density corresponds to $k_F=1.85\ \text{fm}^{-1}$, as listed in Table~\ref{tab1}.

The RBHF results in the full Dirac space by Huber, Weber, and Weigel~\cite{Huber1995_PRC51-1790}, utilizing the techniques of relativistic many-body Green's functions, are shown as the blue dashed line. 
They agree with our results below the saturation density.
The discrepancy above the saturation density is found mainly arising from the different schemes for the starting energies $W$ in Eqs.~\eqref{20200218-eq4} and \eqref{20200218-eq5}.

In this work the NESs in Eqs.~\eqref{20200218-eq4} and \eqref{20200218-eq5} are treated as occupied states.
In the relativistic Green's function approach, the quantities corresponding to the starting energies in Eqs.~\eqref{20200218-eq3}, \eqref{20200218-eq4}, and \eqref{20200218-eq5} are all chosen as $W=E^{+}_{\bm{p}}+E^{+}_{\bm{p}'}$. 
This scheme is the same as the one to treat the NESs as unoccupied states in the Brueckner theory. 
We also performed the calculation with this unoccupied choice, and the discrepancy above the saturation density could be eliminated to a large extent. 
To our knowledge, although there are some discussions on the similarities and differences between the Brueckner theory and the Green's function approach~\cite{Haensel1978_ZPA284-83,Weber1985_PRC32-2141}, a comprehensive comparison has, so far, not yet been carried out~\cite{Crichton1973_APNY75-77}.

In Fig.~\ref{Fig1}, we also display as the green dotted line for the RBHF results in the full Dirac space by Katayama and Saito~\cite{Katayama2015_PLB747-43,Katayama2014_arXiv1410.7166}. 
Their results are different from the ones in this work and Ref.~\cite{Huber1995_PRC51-1790}.
In contrast to the c.m. frame adopted in the most RBHF calculations, in Refs.~\cite{Katayama2015_PLB747-43,Katayama2014_arXiv1410.7166} the matrix elements of the Bonn potential are calculated in the rest frame.
However, we have shown that the calculations of the matrix elements of $NN$ interaction in the rest frame and in the c.m. frame lead to only minor differences, by analyzing the effect of the approximation in Eq.~\eqref{eq:transf}. 
We also notice that in Ref.~\cite{Katayama2014_arXiv1410.7166}, the binding energy per nucleon calculated with Eq.~\eqref{method2} by using the single-particle potentials, does not agree with the one obtained with Eq.~\eqref{20210112-eq1}.

\begin{figure}[htbp]
  \centering
  \includegraphics[width=10.0cm]{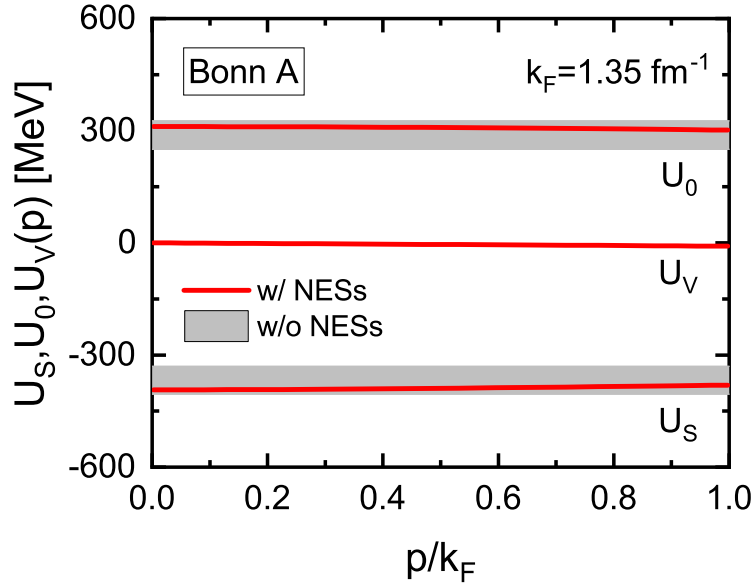}
  \caption{(Color online) Momentum dependence of the single-particle potentials $U_S$, $U_0$, and $U_V$ calculated by the RBHF theory with NESs (red solid line) at $k_F=1.35\ \text{fm}^{-1}$ with the potential Bonn A~\cite{Machleidt1989_ANP19-189}.
  The shaded areas denote the uncertainties of the RBHF results without NESs (see text for more details).}
  \label{Fig2}
\end{figure}

Fig.~\ref{Fig2} presents the single-particle potentials $U_S$, $U_0$, and $U_V$ at $k_F=1.35\ \text{fm}^{-1}$ calculated as functions of momentum. Obviously these quantities exhibit only a weak momentum dependence.
In addition, the strength of the spacelike component of the vector potential $U_V$ is extremely small as compared to the remaining two components.

To investigate the uncertainties of single-particle potentials obtained in the Dirac space with PESs only, we use the momentum-independence approximation introduced in Ref.~\cite{Brockmann1990_PRC42-1965}.
$U_V$ is neglected and the momentum-independent potentials $U_S$ and $U_0$ are extracted from the single-particle potential energies at two momenta. By varying in this procedure one momentum from $0.1\ k_F$ to $0.9\ k_F$ and the other from $0.2\ k_F$ to $1.0\ k_F$ in steps of $0.1\ k_F$, the RBHF calculations in the Dirac space with the PESs only are performed and the uncertainties are shown schematically by the shaded regions in Fig.~\ref{Fig2}.
Considerable uncertainties of about $70$ MeV are found for both potentials $U_S$ and $U_0$. These results again demonstrate the importance of the calculations in the full Dirac space.

\begin{figure}[htbp]
  \centering
  \includegraphics[width=8.0cm]{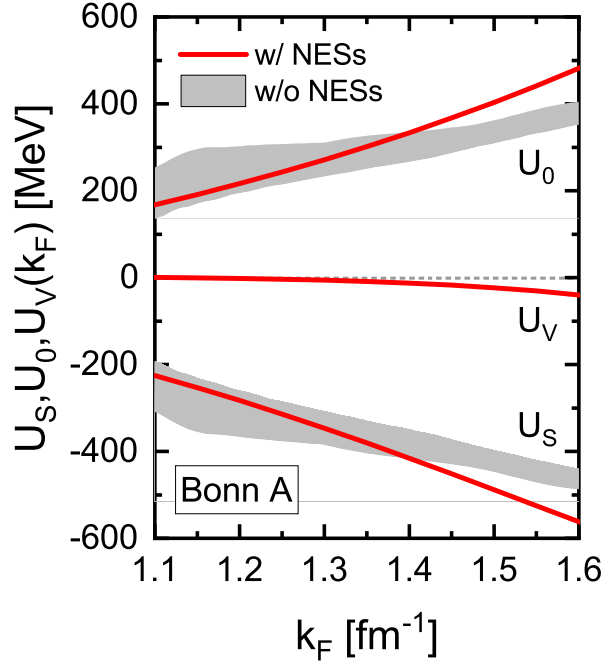}
  \caption{(Color online) Density dependence of the single-particle potentials $U_S$, $U_0$, and $U_V$ at the Fermi momentum calculated by the RBHF theory with NESs (red solid line) with the potential Bonn A~\cite{Machleidt1989_ANP19-189}.
  The shaded areas denote the uncertainties of the RBHF calculations without NESs (see text for more details).}
  \label{Fig3}
\end{figure}

In Fig.~\ref{Fig3}, the single-particle potentials $U_S$, $U_0$ and $U_V$ at the Fermi momentum calculated in the full Dirac space are shown as functions of the Fermi momentum $k_F$. A strong density dependence is found for $U_S$ and $U_0$, while it is less pronounced for $U_V$. We also show in Fig.~\ref{Fig3} the results obtained by the RBHF calculations without NESs. For $U_S$ and $U_0$ at $k_F=1.14\ \text{fm}^{-1}$ we found uncertainties up to $126$ and $128$ MeV, respectively. Although these uncertainties are reduced with the increasing density, the strengths of $U_S$ and $U_0$ are both underestimated above the saturation density, as compared to the results obtained in the full Dirac space.

\begin{figure}[htbp]
  \centering
  \includegraphics[width=9.0cm]{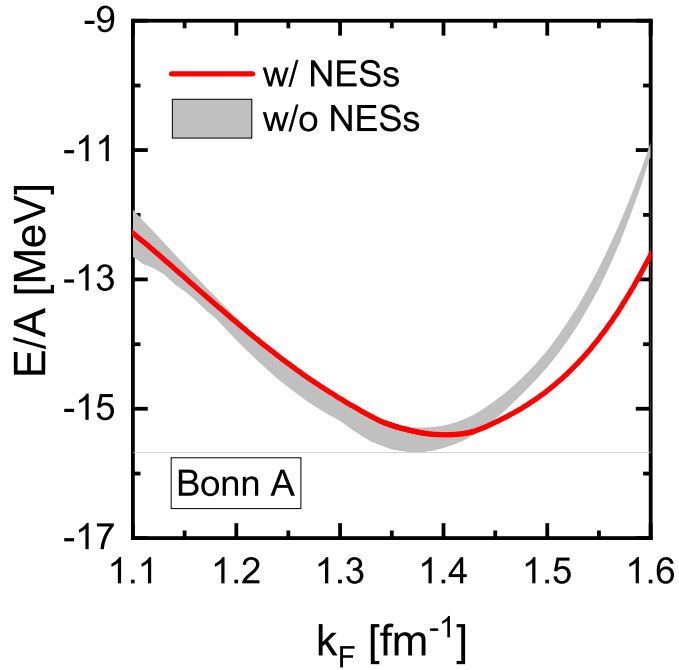}
  \caption{(Color online) Binding energy per nucleon $E/A$ of symmetric nuclear matter as a function of the Fermi momentum $k_F$ calculated by the RBHF theory in the full Dirac space  (red solid line) using the potential Bonn A~\cite{Machleidt1989_ANP19-189}.
  The shaded area denotes the uncertainties of the RBHF calculations without NESs (see text for more details).}
  \label{Fig4}
\end{figure}

Since the binding energy of nuclear matter generally results from a sensitive cancellation between single-particle potentials, it is interesting and necessary to study also the uncertainties of the equation of state in the Dirac space with PESs only. In Fig.~\ref{Fig4} we show the binding energy per nucleon $E/A$ as a function of the Fermi momentum $k_F$ calculated without NESs, in comparison with the one calculated in the full Dirac space.
It is found that the uncertainties of the binding energy per nucleon can reach $0.7$ MeV at $k_F=1.10\ \text{fm}^{-1}$.
Above the saturation density, less binding is obtained. Moreover, for the calculations in the Dirac space with PESs only, the compression modulus $K_\infty$ ranges from $218$ to $426$ MeV.
Again, this shows the importance to perform the RBHF calculations in the full Dirac space.

In Table~\ref{tab1}, we summarize the saturation properties of nuclear matter obtained in this work by solving the RBHF equations in the full Dirac space with the potentials Bonn A, B, and C. Corresponding non-relativistic BHF results, successful phenomenological covariant density functionals~\cite{Meng2016Book} NL3~\cite{Lalazissis1997_PRC55-540}, DD-ME2~\cite{Lalazissis2005_PRC71-024312}, DD-PC1~\cite{Niksic2008_PRC78-034318}, PC-PK1~\cite{ZHAO-PW2010_PRC82-054319}, and PKO1~\cite{Long2006_PLB640-150} as well as empirical values are listed for comparison. 
The binding energy per nucleon obtained by the RBHF calculations in the full Dirac space with Bonn A is $-15.40$ MeV, which is in agreement with the empirical values $-16\pm1$ MeV, and the saturation density $\rho$ is $0.188\ \text{fm}^{-3}$, which is slightly higher than the empirical ones $0.16\pm0.01\ \text{fm}^{-3}$.
For Bonn B, the saturation density is described satisfactorily, while the binding energy is inadequate. Bonn C leads to smaller values for both the density and the binding energy at the saturation density. This trend for Bonn A, B and C is similar to the case found in Ref.~\cite{Brockmann1990_PRC42-1965}.
The compression modulus of nuclear matter at the saturation density is 258 MeV for Bonn A, which is in good agreement with the empirical values of $240\pm20$ MeV~\cite{Garg2018_PPNP101-55}.
In the last column, the Dirac masses at the Fermi momentum in unit of nucleon mass are also shown, which are close to the values in phenomenological covariant energy density functionals.

\begin{table}[htbp]
\caption{
Saturation properties of symmetric nuclear matter calculated by the RBHF theory in the full Dirac space using the bare $NN$ interactions Bonn A, B and C~\cite{Machleidt1989_ANP19-189}:
the saturation density $\rho_0$, the binding energy per nucleon $E/A$, the compression modulus $K_\infty$, and the Dirac mass $M_D^*/M$ at the saturation density. They are compared with the corresponding values from the non-relativistic BHF calculations and the phenomenological covariant density functionals NL3~\cite{Lalazissis1997_PRC55-540}, DD-ME2~\cite{Lalazissis2005_PRC71-024312}, DD-PC1~\cite{Niksic2008_PRC78-034318}, PC-PK1~\cite{ZHAO-PW2010_PRC82-054319}, and PKO1~\cite{Long2006_PLB640-150}.
The empirical values are listed in the last row.}
\begin{tabular}{ccccccccc}
\hline
\hline
Potential  & &  $\rho_0\ [\text{fm}^{-3}]$  & &  $E/A\ [\text{MeV}]$  & & $K_{\infty}\ [\text{MeV}]$  & & $M_D^*/M$ \\ [0pt]
\hline
RBHF Bonn A  & $\qquad$  &  0.188   & $\qquad$    &  -15.40 & $\qquad$  &  258 & $\qquad$ & 0.55 \\ [-4pt]
RBHF Bonn B  & $\qquad$  &  0.164   & $\qquad$    &  -13.36 & $\qquad$  &  206 & $\qquad$ & 0.61 \\ [-4pt]
RBHF Bonn C  & $\qquad$  &  0.144   & $\qquad$    &  -12.09 & $\qquad$  &  150 & $\qquad$ & 0.65 \\ [-4pt]
BHF Bonn A   & $\qquad$  &  0.428   & $\qquad$    &  -23.55 & $\qquad$  &  204 & \\ [-4pt]
BHF Bonn B   & $\qquad$  &  0.309   & $\qquad$    &  -18.30 & $\qquad$  &  160 & \\ [-4pt]
BHF Bonn C   & $\qquad$  &  0.247   & $\qquad$    &  -15.75 & $\qquad$  &  103 & \\ [-4pt]
NL3        & $\qquad$  &   0.148 & $\qquad$    &  -16.30 & $\qquad$  &  272 & $\qquad$ & 0.60 \\ [-4pt]
DD-ME2     & $\qquad$  &   0.152 & $\qquad$    &  -16.14 & $\qquad$  &  251 & $\qquad$ & 0.57 \\ [-4pt]
DD-PC1     & $\qquad$  &   0.152 & $\qquad$    &  -16.06 & $\qquad$  &  230 & $\qquad$ & 0.58 \\ [-4pt]
PC-PK1     & $\qquad$  &   0.154 & $\qquad$    &  -16.12 & $\qquad$  &  238 & $\qquad$ & 0.59 \\ [-4pt]
PKO1       & $\qquad$  &   0.152 & $\qquad$    &  -16.00 & $\qquad$  &  250 & $\qquad$ & 0.59 \\ [-4pt]
Empirical  & $\qquad$  &   0.16 $\pm0.01$  & $\qquad$    &  -16$\pm1$     & $\qquad$  &  240$\pm20$ & $\qquad$ &   \\ [0pt]
\hline
\hline
\end{tabular}
\label{tab1}
\end{table}

\newpage

\section{Summary}\label{SecV}

In summary, the RBHF equations have been solved for symmetric nuclear matter in the full Dirac space with the Bonn potential.
In this way the uncertainties in the RBHF calculations in the Dirac space with PESs only could be avoided. 
The Thompson equation is chosen as the scattering equation and the matrix elements of the Bonn potential are treated in the c.m. frame. 
The obtained saturation properties of symmetric nuclear matter are in good agreement with the empirical values.
The equation of state agrees with the results based on the relativistic Green's function approach up to the saturation density.
The discrepancy above the saturation density is found mainly arising from the different schemes for the starting energies.
Uncertainties of the RBHF calculation in the Dirac space with PESs only have been analyzed.
It is found that the uncertainties of the single-particle potentials can reach more than $100$ MeV, and the equation of state is less bound above the saturation density. These analyses demonstrate the significance of the RBHF calculations in the full Dirac space.

\begin{acknowledgments}
Sibo Wang thanks Pengwei Zhao, Xiulei Ren, Shihang Shen, Hui Tong and Weijiang Zou for helpful discussions. This work was supported in part by the National Key R\&D Program of China (No. 2017YFE0116700 and 2018YFA0404400), the National Natural Science Foundation of China (No. 11935003, No. 11975031, No. 11875075 and No. 12070131001), and the DFG cluster of excellence "Origin and Structure of the Universe".
Part of this work was achieved by using the High-performance Computing Platform of Peking University, and the supercomputer OCTOPUS at the Cybermedia Center, Osaka University under the support of Research Center for Nuclear Physics of Osaka University.
\end{acknowledgments}

\appendix

\section{Partial-wave decomposition of the Thompson equation}\label{App-ThEqu}

In this appendix we give the details for the solution of the Thompson equation~\eqref{eqThomJ} in the full Dirac space with partial-wave decomposition.
There are $2^4=16$ possible combinations of PESs and NESs for the bare $NN$ interaction such as $V^{-+++}$.
The $\pm$-signs in the superscript are used to denote the PESs or NESs.
Due to the \emph{no-sea} approximation and the limitation of intermediate states as PESs in the Thompson equation~\cite{Thompson1970_PRD1-110},
four combinations $V^{++++}, V^{-+++}, V^{-+-+}$ and $V^{-++-}$ are needed in practice.
For a given combination, the number of independent helicity amplitudes is $2^4=16$ for each partial wave.
This number can be reduced to eight independent amplitudes due to the symmetries under parity transformation
\beq
  \begin{split}
  	\la \lambda'_1\lambda'_2|V_{J}^{++++}(q',q)|\lambda_1\lambda_2\ra
   =&\ + \la -\lambda'_1-\lambda'_2|V_{J}^{++++}(q',q) |-\lambda_1-\lambda_2\ra,  \\
   \la \lambda'_1\lambda'_2|V_{J}^{-+++}(q',q)|\lambda_1\lambda_2\ra
   =&\ - \la -\lambda'_1-\lambda'_2|V_{J}^{-+++}(q',q) |-\lambda_1-\lambda_2\ra, \\
   \la \lambda'_1\lambda'_2|V_{J}^{-+-+}(q',q)|\lambda_1\lambda_2\ra
   =&\ + \la -\lambda'_1-\lambda'_2|V_{J}^{-+-+}(q',q) |-\lambda_1-\lambda_2\ra,  \\
   \la \lambda'_1\lambda'_2|V_{J}^{-++-}(q',q)|\lambda_1\lambda_2\ra
   =&\ + \la -\lambda'_1-\lambda'_2|V_{J}^{-++-}(q',q) |-\lambda_1-\lambda_2\ra,
  \end{split}
\eeq
where $J$ is the total angular momentum for each partial wave. The magnitudes of the relative momenta of the two nucleons in the initial and final states are denoted by $q$ and $q'$. $\lambda_i$ and $\lambda'_i\ (i=1,2)$ represent the helicities of the nucleon $i$ in the initial and final states, respectively.

The eight independent helicity amplitudes are chosen as follows:
\beq
  \begin{split}
    &V_1^{J}(q',q) \equiv\la++|V^{J}(q',q) |++{\ra},\\
    &V_2^{J}(q',q) \equiv\la++|V^{J}(q',q) |--{\ra},\\
    &V_3^{J}(q',q) \equiv\la+-|V^{J}(q',q) |+-{\ra},\\
    &V_4^{J}(q',q) \equiv\la+-|V^{J}(q',q) |-+{\ra},\\
    &V_5^{J}(q',q) \equiv\la++|V^{J}(q',q) |+-{\ra},\\
    &V_6^{J}(q',q) \equiv\la+-|V^{J}(q',q) |++{\ra},\\
    &V_7^{J}(q',q) \equiv\la++|V^{J}(q',q) |-+{\ra},\\
    &V_8^{J}(q',q) \equiv\la+-|V^{J}(q',q) |--{\ra},
  \end{split}
\eeq
where the $\pm$-signs correspond to the signs of the helicities and the labels of PESs and NESs have been suppressed. To partially decouple this system, it is useful to introduce the following linear combinations of helicity amplitudes:
\beq
  \begin{aligned}
    {}^0V^J\    \equiv\ & \ V_{1}^J-V_{2}^J,\\
    {}^1V^J\    \equiv\ & \ V_{3}^J-V_{4}^J,\\
    {}^{12}V^J\ \equiv\ & \ V_{1}^J+V_{2}^J,\\
    {}^{34}V^J\ \equiv\ & \ V_{3}^J+V_{4}^J,\\
    {}^{57}V^J\ \equiv\ & \ V_{5}^J+V_{7}^J,\\
    {}^{68}V^J\ \equiv\ & \ V_{6}^J+V_{8}^J,\\
    {}^{2}V^J\  \equiv\ & \ V_{5}^J-V_{7}^J,\\
    {}^{3}V^J\  \equiv\ & \ V_{6}^J-V_{8}^J.
  \end{aligned}
\eeq
Corresponding definitions for $G^J$ are also introduced. Using these definitions, the system of Thompson equation for a given combination can be partially decoupled.
Taking $V^{-+++}_J$ and $G^{-+++}_J$ as an example, two subsets of coupled integral equations are obtained
\beq\label{eq:Gset1}
\begin{split}
  {}^0G^{-+++}_J
    =&\ {}^0V^{-+++}_J + \int\frac{M_{\text{av}}^{*2} }{E_{\text{av}}^{*2}}
    \frac{Q_{\mathrm{av}}}{W-2E_{\text{av}}}
      \left[ {}^0V^{-+++}_J \cdot {}^0 G^{++++}_J + {}^2V^{-+++}_J \cdot {}^3G^{++++}_J  \right], \\
  {}^1G^{-+++}_J
    =&\ {}^1V^{-+++}_J + \int\frac{M_{\text{av}}^{*2} }{E_{\text{av}}^{*2}}
      \frac{Q_{\mathrm{av}}}{W-2E_{\text{av}}}
      \left[ {}^3V^{-+++}_J \cdot {}^2 G^{++++}_J + {}^1V^{-+++}_J \cdot {}^1G^{++++}_J  \right], \\
  {}^2G^{-+++}_J
    =&\ {}^2V^{-+++}_J + \int\frac{M_{\text{av}}^{*2} }{E_{\text{av}}^{*2}}
       \frac{Q_{\mathrm{av}}}{W-2E_{\text{av}}}
      \left[ {}^0V^{-+++}_J \cdot {}^2 G^{++++}_J + {}^2V^{-+++}_J \cdot {}^1G^{++++}_J  \right], \\
  {}^3G^{-+++}_J
    =&\ {}^3V^{-+++}_J + \int\frac{M_{\text{av}}^{*2} }{E_{\text{av}}^{*2}}
     \frac{Q_{\mathrm{av}}}{W-2E_{\text{av}}}
      \left[ {}^3V^{-+++}_J \cdot {}^0 G^{++++}_J + {}^1V^{-+++}_J \cdot {}^3G^{++++}_J  \right], \\
\end{split}
\eeq
and
\beq\label{eq:Gset2}
\begin{split}
  {}^{12}G^{-+++}_J
    =&\ {}^{12}V^{-+++}_J + \int  \frac{M_{\text{av}}^{*2} }{E_{\text{av}}^{*2}}  \frac{Q_{\mathrm{av}}}{W-2E_{\text{av}}}
      \left[ {}^{12}V^{-+++}_J \cdot {}^{12} G^{++++}_J + {}^{57}V^{-+++}_J \cdot {}^{68}G^{++++}_J  \right], \\
  {}^{34}G^{-+++}_J
    =&\ {}^{34}V^{-+++}_J + \int  \frac{M_{\text{av}}^{*2} }{E_{\text{av}}^{*2}}  \frac{Q_{\mathrm{av}}}{W-2E_{\text{av}}}
      \left[ {}^{68}V^{-+++}_J \cdot {}^{57} G^{++++}_J + {}^{34}V^{-+++}_J \cdot {}^{34}G^{++++}_J  \right], \\
  {}^{57}G^{-+++}_J
    =&\ {}^{57}V^{-+++}_J + \int  \frac{M_{\text{av}}^{*2} }{E_{\text{av}}^{*2}}  \frac{Q_{\mathrm{av}}}{W-2E_{\text{av}}}
      \left[ {}^{12}V^{-+++}_J \cdot {}^{57} G^{++++}_J + {}^{57}V^{-+++}_J \cdot {}^{34}G^{++++}_J  \right], \\
  {}^{68}G^{-+++}_J
    =&\ {}^{68}V^{-+++}_J + \int  \frac{M_{\text{av}}^{*2} }{E_{\text{av}}^{*2}}  \frac{Q_{\mathrm{av}}}{W-2E_{\text{av}}}
      \left[ {}^{68}V^{-+++}_J \cdot {}^{12} G^{++++}_J + {}^{34}V^{-+++}_J \cdot {}^{68}G^{++++}_J  \right]. \\
\end{split}
\eeq
The coupled integral equations for other combinations $G^{++++}_J, G^{-+-+}_J, G^{-++-}_J$ can be obtained in a complete analogy to Eqs.~\eqref{eq:Gset1} and \eqref{eq:Gset2}.
These matrix equations can be solved with the standard method of Haftel and Tabakin~\cite{Haftel1970_NPA158-1}.

\section{The Calculations of the Matrix Elements of the Single-particle Potential Operator}\label{App-Sgm}

The matrix elements of the single-particle potential operator in Eq.~\eqref{20200515-eq1} can be calculated with the $G$ matrix coupled to the total angular momentum in the helicity scheme. The transformations for $\Sigma^{++}$ and $\Sigma^{--}$ are trivial, which are given by~\cite{Erkelenz1971_NPA176-413}
\beq\label{20200510-eq11}
	\Sigma^{++}(p) = \int^{k_F}_0 \frac{d^3p'}{(2\pi)^3} \frac{M^*_{\bm{p}'}}{E^*_{\bm{p}'}}
      \frac{1}{4} \sum_{JT\lambda_1\lambda_2} \frac{(2J+1)(2T+1)}{4\pi}
      \la \lambda_1\lambda_2|G^{++++}_{JT}(q,q|P,W)(1-P_{12})|\lambda_1\lambda_2 \ra ,\\   
\eeq
and 
\beq\label{20200515-eq21}    
  \Sigma^{--}(p) = \int^{k_F}_0 \frac{d^3p'}{(2\pi)^3} \frac{M^*_{\bm{p}'}}{E^*_{\bm{p}'}}
      \frac{1}{4} \sum_{JT\lambda_1\lambda_2} \frac{(2J+1)(2T+1)}{4\pi}
      \la \lambda_1\lambda_2|G^{-+-+}_{JT}(q,q|P,W)(1-P_{12})|\lambda_1\lambda_2 \ra. 
\eeq
where $T$ is the total isospin and $P_{12}$ is the exchange operator.
In our calculation, the amplitude $P$ is calculated as $P=|\bm{P}| = |\bm{p}+\bm{p}'|/2$.
$q$ is the amplitude of $\bm{q}$ in the c.m. frame, which is obtained with a strict Lorentz transformation from $\bm{p}$ in the rest frame, as in Ref.~\cite{Horowitz1987_NPA464-613}.
Taking a four-vector $a^\mu=(a^0,\bm{a})$ in the rest frame of nuclear matter with non-vanishing $\bm{P}$ as an example, the corresponding four-vector $a_c^\mu=(a_c^0,\bm{a}_c)$ in the c.m. frame where $\bm{P}=0$ is determined by~\cite{Horowitz1987_NPA464-613}
\begin{subequations}
  \begin{align}
   \bm{a}_{c} &= \bm{a} + \bm{\beta} \gamma\left(\frac{\gamma}{\gamma+1} \bm{\beta} \cdot \bm{a}-a^{0}\right),\\
   a_{c}^{0} &= \gamma\left(a^{0}-\bm{\beta} \cdot \bm{a}\right).
  \end{align}
\end{subequations}
The parameters $\bm{\beta}$ and $\gamma$ are found as $\bm{\beta}= \bm{P}/P^{0}, \gamma=\left(1-\bm{\beta}^{2}\right)^{-1 / 2}$, where $P^0 = (E_{\bm{p}} + E_{\bm{p}'})/2$.

$\Sigma^{-+}$ is more complicated. With explicit spin and isospin degrees of freedom, $\Sigma^{-+}$ is calculated as
\beq\label{20201027-eq21}
  \Sigma^{-+}(p) = \sum_{s_2t_2} \int \frac{d^3p'}{(2\pi)^3} \frac{M^*_{\bm{p}'}}{E^*_{\bm{p}'}}
  \la s_1s_2t_1t_2|G^{-+++}(\bm{q},\bm{q}|\bm{P},W)(1-P_{12})| s_1s_2t_1t_2\ra,
\eeq
with $s_1=\frac{1}{2}$. Following Appendix B of Ref.~\cite{Anastasio1981_PRC23-2273},
the $G$ matrix must be expressed in the angular momentum projected helicity basis as follows
\beq\label{20200219-eq1}
  \begin{split}
    &\sum_{s_2t_2}\la s_1s_2t_1t_2|G^{-+++}(\bm{q},\bm{q}|\bm{P},W)(1-P_{12})| s_1s_2t_1t_2\ra \\
    =&\ \frac{1}{2} \sum_{LL'JTSS'\mathcal{L}\lambda_1\lambda_2\lambda'_1\lambda'_2 m}
    (-1)^{J+L+L'+S+S'+s_1-\frac{1}{2}}
    \frac{(2J+1)(2T+1)[(2S+1)(2S'+1)(2L+1)(2L'+1)]^{1/2}}{4\pi} \\
    &\times \la \lambda'_1\lambda'_2 |G^{-+++}_{JT}(q,q|P,W)(1-P_{12})| \lambda_1\lambda_2 \ra
    \la L'S'J|\lambda'_1\lambda'_2 J \ra
    \la \lambda_1\lambda_2 J|LSJ\ra \\
    &\times \bpm  L & L' & \mathcal{L} \\ 0 & 0 & 0\epm
    \bpm \frac{1}{2} & \frac{1}{2} & \mathcal{L} \\ -s_1 & s_1 & m\epm
    \bBm \mathcal{L} & S' & S \\ J & L & L'\eBm
    \bBm \mathcal{L} & \frac{1}{2} & \frac{1}{2} \\ \frac{1}{2} & S & S'\eBm
    \sqrt{4\pi (2\mathcal{L}+1)}
    Y^*_{\mathcal{L}m}(\hat{\bm{q}}),
  \end{split}
\eeq
where
\beq
  \la LSJ|\lambda_1\lambda_2J\ra
  = \left(\frac{2L+1}{2J+1}\right)^{1/2} \left( L0S\lambda_1 - \lambda_2 | J \lambda_1 - \lambda_2 \right) 
    \left( \frac{1}{2}\lambda_1\frac{1}{2}-\lambda_2 | S\lambda_1-\lambda_2\right).
\eeq
Terms in curly brackets are $6j$-symbols in quantum angular momentum theory~\cite{Varshalovich1988}.


%

\end{document}